\documentclass[12pt]{article}
 \usepackage{amssymb}

 \newcommand{\Z}{ {\mathbb Z} }
 \newcommand{\fnm}{\footnotemark}
 \newcommand{\fnt}{\footnotetext}

\begin{document}

 \vspace{15pt}

 \begin{center}
 \large\bf

  On black holes in
  multidimensional theory with  Ricci-flat  internal spaces

 \vspace{15pt}

 \normalsize\bf
  S.B. Fadeev, V.D. Ivashchuk and V.N. Melnikov

 \vspace{7pt}

 \it  \ \ \ Center for Gravitation and Fundamental
 Metrology,  VNIIMS, 46 Ozyornaya Str., Moscow 119361, Russia
 \fnm[1]\fnt[1]{the current address}  \\

 \end{center}
 \vspace{15pt}

  \begin{abstract}
 A generalization of the Tangherlini solution
 for the case of $n$ internal Ricci-flat spaces is obtained. It is shown
 that in the $(2+d)$-dimensional section  a horizon
 exists only in the trivial case when the internal-space factors are constant.
 The $p$-adic analog  of the solution is also considered.
 \end{abstract}

 \setcounter{equation}{0}

{\bf 1.} In ref. \cite{BIM-89}  the Schwarzschild solution is
generalized for the case of $n$ internal Ricci-flat spaces. It was
shown that a horizon in the four-dimensional section of the metric
exists only when the internal space scale factors are constant.
This proposition was also suggested in ref. \cite{CMP}, where a
special case of the solution \cite{BIM-89}  with $n =2$ internal
spaces (one of which being one dimensional) was considered.

In this paper we consider exact static, spherically symmetric
solutions of the Einstein equations in $(2+d+N_1+ \ldots
 + N_n)$-dimensional gravity $(d\geq 2)$ with a chain of $n$
Ricci-flat internal spaces. We show that as in the $d=2$ case,
considered in ref. \cite{BIM-89}, a horizon is absent in all
nontrivial cases. Finally, we consider a formal analog of the
solution for the case of $p$-adic numbers \cite{Koblitz}.

{\bf 2.} We consider the Einstein equations

\begin{equation}
 R_{MN}=0  \label{1}
\end{equation}

on the $D$-dimensional manifold

 \begin{equation}
 M=M_0\times M_1\times \ldots \times M_n, \label{2}
 \end{equation}

where
\[
 \dim \ M_i=N_i, \ \ D=2+d+\sum_{j=1}^{n}N_j, \ \ i>0,
\]
$M_0$ is $(2+d)$-dimensional space-time $(d\geq 2)$ and $M_i$ are
Ricci-flat manifolds with the metrics $g_{(i)}, i=1, \dots, n$. We
seek solutions of (1) such that $M_0$ is static, spherically
symmetric $(O(d+1)$-symmetric), while all the scale factors
$\exp(\beta_i)$ of the internal spaces $M_i$ depend on the radial
coordinate $u$, i.e., the $D$-metric is
\begin{eqnarray}
 g=&-& \exp[2\gamma (u)] dt\otimes dt + \exp[2\alpha (u)]du \otimes du \nonumber \\
  &+& \exp[2\beta (u)] d\Omega_d^2 + \sum_{i=1}^{n} \exp[2\beta_i(u)]g_{(i)},
   \label{3}
\end{eqnarray}
where $d\Omega_d^2=g_{(0)}$ is the standard $S^d$ metric.

If we denote $\gamma =\beta_{-1},N_{-1}=1$ and $\beta =\beta_0,
 N_0=d$ and choose the harmonic radial coordinate $u$ such that
 $\alpha =\sum_{k=-1}^{n}\beta_iN_i$ then the Einstein equations
(\ref{1}) can be written in the form
\begin{eqnarray}
R_{11}&=&\sum_{i=-1}^{n}(-\beta_i^{''}+ \alpha^{'}\beta_i^{'}-
(\beta_i^{'})^2) N_i=0, \nonumber \\
R_{00} &=& \exp(2\beta_{-1}-2\alpha )\beta_{-1}^{''}=0, \nonumber \\
R_{k l} &=& g_{(0)k l}[d-1-\beta_0^{''} \exp(2\beta_0 -2\alpha)]=0. \nonumber \\
R_{m_i n_i} &=& -g_{(i)m_in_i}\beta_i^{''} \exp(2\beta_i-2\alpha),
\ i=1,\ldots ,n.  \label{4}
\end{eqnarray}

This set of equations is easily solved, so that the metric $g$
(\ref{3}) after an appropriate redefinition of the radial
coordinate $(u\Rightarrow R=R(u))$ may be written in the following
way,

\begin{eqnarray}
 g&=& -c^2[1-\varepsilon (L/R)^{d-1}]^adt\otimes dt \nonumber \\
 &+& [1-\varepsilon (L/R)^{d-1}]^{(a+b+d-2)/(1-d)}dR\otimes dR \nonumber \\
 &+& [1-\varepsilon (L/R)^{d-1}]^{(a+b-1)/(1-d)}R^2 d\Omega_d^2 \nonumber \\
 &+& \sum_{i=1}^{n} c_i[1-\varepsilon (L/R)^{d-1}]^{a_i}g_{(i)}, \label{5} \\
 \varepsilon &=& \pm 1, \nonumber
 \end{eqnarray}
where $L\geq 0$, $R>0$ and $R>L$ for $\varepsilon =+1$; $L$,
 $c \neq 0$ and $c_i\neq 0$ are constants,

\begin{equation}
 b = \sum_{i=1}^{n}a_iN_i, \label{6}
\end{equation}
and the constants $a,a_1,\dots ,a_n$ satisfy the relation

\begin{equation}
 (a+b)^2+(d-1)\left(a^2+\sum_{i=1}^{n}a_i^2N_i\right)=d. \label{7}
\end{equation}

In the case $d=2$ and $\varepsilon =+1$ this solution coincides
with that of ref. \cite{BIM-89}. In the special case $n=1$ and
  \[
  a = \left(\frac{d(d + N_1 -1)}{(d-1)(d+ N_1)} \right)^{1/2},
  \qquad a_1 = - \frac{a}{d+ N_1 - 1}, \nonumber
  \]
 this solution was considered earlier in ref. \cite{Myers}. The case $d=2$,
 $n=1$ was considered in ref. \cite{Yoshimura}, and its scalar-vacuum
 generalization was obtained in ref. \cite{BrIv-89}.

 {\bf 3}. Let us consider the $(2+d)$-dimensional section of the metric
 (\ref{5}). In the case $L=0$ the metric is flat, while for $L>0$ and

 \begin{equation}
 a-1=a_1= \ldots =a_n=0 \label{8}
 \end{equation}
 it coincides with the Tangherlini solution \cite{Tangherlini}.

Now let us prove that a horizon at $R=L$ ($L>0$) takes place only
in the case (\ref{8}) for $\varepsilon =+1$. Indeed, for the light
propagating along a radius from a place with $R=R_0$ towards the
center the coordinate time interval is
 \begin{equation}
 t-t_0=\frac{1}{c}\displaystyle\int_{R}^{R_0}dx[1-(L/R)^{d-1}]^{\lambda},
 \label{9}
 \end{equation}
where
 \begin{equation}
 \lambda =\frac{1}{2}\left(\frac{a+b+d-2}{1-d}-a\right). \label{10}
 \end{equation}
Relation (\ref{7}) is equivalent to the identity
 \begin{equation}
 (a+b/d)^2 =
 1-\frac{d-1}{d}\sum_{i=1}^{n}a_i^2N_i-\frac{b^2}{d^2}(d-1). \label{11}
 \end{equation}
Let $\varepsilon=+1$. If some $a_i$ ($i=1,\dots,n$) are nonzero,
then by (\ref{11}) $a + b/d<1$ and from (\ref{10}) $\lambda >-1$,
hence, the integral (\ref{9}) converges at $R=L$. This means that
a radial light beam reaches the surface $R=L$ in a finite time
interval, i.e. it is not a horizon. When $a_i=0$, $i=1,\dots ,n$,
then $a=\pm 1$. For the Tangherlini case $a=+1$ we have a horizon
$(\lambda =-1)$, and for $a=-1$ ($\lambda =1/(d-1)$) the horizon
at $R=L$ is absent. Evidently, for $\varepsilon =-1$ the horizon
at $R=L$ is absent too. This completes the proof.

{\bf 4}. At present there is more interest in considering the
physical models with $p$-adic numbers \cite{Koblitz} instead of
real ones. This interest was stimulated mainly by the pioneering
works on $p$-adic strings  \cite{Volovich,FreundOlson}. Recently a
$p$-adic generalization of the classical and quantum gravitational
theory was defined \cite{ADFV} and some solutions of the Einstein
equations were considered \cite{ADFV,IMF-90-Yak}. In this section
we consider the $p$-adic analog of the solution (\ref{5}). Let us
briefly recall the definition of $p$-adic numbers
\cite{Koblitz,Mahler}. Let $p$ be a prime number. Any rational
number $a\neq 0$ can be represented in the form $a=p^km/n$, where
the integer numbers $m$ and $n$ are not divisible by $p$. Then the
$p$-adic norm is defined as follows: $\mid a\mid_p=p^{-k}$. This
norm is non-Archimedean: $\mid a+b \mid_p \leq \max (\mid a
\mid_p,\mid b \mid_p)$. The completion of $Q$ with this norm is
the $p$-adic number field $Q_p$. Any nonzero $p$-adic number $a
\in Q_p$ can be uniquely represented as the series
\[
 a=p^k(a_0+a_1p+a_2p^2+ \ldots ), \nonumber
\]
where $a_0=1,\dots ,p-1$, and $a_i=0, \dots ,p-1$ for $i>0$.

The definitions of derivatives, manifold and tensor analysis in
the $p$-adic case are similar to those of the real case. The power
$p$-adic function is defined as follows,
\begin{equation}
(1+x)_p^{\alpha} \equiv \exp_p\{\alpha [\log_p(1+x)]\}, \label{12}
\end{equation}
where $\mid x\mid_p<1$ and $\mid \alpha \mid_p \mid
 x \mid_p< \delta_p$. Here $\delta_p=1$ for $p\neq 2$ and
$\delta_2=\frac{1}{2}$. The definition is correct, for the
functions $\exp_p$ and $\log_p$ are well defined on the discs
$\{\mid x \mid_p < \delta_p\}$ and $\{\mid x-1 \mid_p < 1\}$
respectively \cite{Koblitz}.

Let us consider the $p$-adic manifold
\begin{equation}
Q_p \times Q_p^{*} \times S^d\times M_1 \times \dots \times M_n,
\label{13}
\end{equation}
where $(S^d,g_{(0)})$ is a space of constant curvature
\[
R_{i j k l}^{0} =
g_{ik}^{(0)}g_{jl}^{(0)}-g_{il}^{(0)}g_{jk}^{(0)}
\]
[$Q_p^{*} \subset Q_p$] and $(M_i,g_{(i)})$ are Ricci-flat
manifolds.

For $R\neq 0$ and
\begin{equation}
\left|\frac{L}{R}\right|_{p}^{d-1}<\min (1,1/\mid a_i\mid_p), \ \
i=-2,-1,\dots ,n,  \label{14}
\end{equation}
with
\[
 a_{-2} = \frac{a+b+d-2}{1-d}, \ \ a_{-1} = a, \ \ a_0=\frac{a+b-1}{1-d},
\]
the metric (\ref{5}) on the manifold (\ref{13}) is well defined.
Then the Einstein equations for the metric (\ref{5}), (\ref{14})
on the manifold (\ref{13}) are satisfied identically, when the
parameters $a,a_1, \dots, a_n \in Q_p$ obey the restriction
 (\ref{7}).

This can be easily checked using the identity
\[
[(1+x)^{\alpha}]^{'} = \alpha (1+x)^{\alpha}/(1+x),
\]
 $\mid x \mid_p,\mid \alpha \mid_p \mid x \mid_p<1$ (the verification
of (\ref{1}) in the $p$-adic case is just the same as in the real
one).

In the $d=2$ case this solution was considered earlier in ref.
\cite{IMF-90-Yak}. It was pointed out that there is an infinite
number or rational solutions of (\ref{7}) in this case. For
example, we may consider the set \cite{IMF-90-Yak}

\begin{eqnarray}
a_1 &=& \frac{4k}{N_1(N_1+2)k^2+1}, \ a_i=0, \ i>1, \nonumber \\
a&=& \frac{-2N_1k\pm [k^2N_1(N_1+2)-1]}{k^2N_1(N_1+2)+1}, \ k \in
 \Z . \nonumber
\end{eqnarray}

In the $p$-adic case there exist pseudo-constant functions
$C=C(R)$ such that $C'(R)=0$ but $C(R)$ is not identically
constant \cite{Mahler}. Such functions may be used in
generalization of well-known solutions of differential equations.
In our case there is also a possibility for the constants
$c,c_1,\dots ,c_n$ and $a,a_1,\dots ,a_n$ to be replaced by the
pseudo-constants (of course, the restriction (\ref{14}) should be
preserved).

There is another possibility to generalize the solution (\ref{5}).
We may suppose that the components of the metric $g_{MN}$ belong
to some extension of $Q_p$. It may be the quadratic extension of
$Q_p$ or even $\Omega_p$, which is the completion of the algebraic
closure of $Q_p$ \cite{Koblitz}. In this case the constants in
(\ref{5}) may belong to the extension of $Q_p$.

The solution (\ref{5}) can be also generalized on scalar-vacuum
and electro-vacuum cases. The last generalization for $d=2$ is
considered in ref. \cite{FIM-91-ChPL}.


\begin{thebibliography}{99}

 \bibitem{BIM-89}
  K.A. Bronnikov, V.D. Ivashchuk and V.N. Melnnikov, in: Problems
  of Gravitation, Plenary Reports of 7th. Soviet Conf. on Gravitation
  (ErGU, Erevan, 1989), p. 70.

 \bibitem{CMP}
 C.G. Callan, R.C. Myers and M.J. Perry., Nucl. Phys.
  B 311 (1988) 673.

 \bibitem{Koblitz}
 N. Koblitz, p-adic numbers, p-adic analysis and   zeta-functions
 (Springer, Berlin, 1977).

 \bibitem{Myers}
 R.C. Myers, Phys. Rev.  D 35 (1987) 455.

 \bibitem{Yoshimura}
 M. Yoshimura, Phys. Rev.  D 34 (1986) 1021.

 \bibitem{BrIv-89}
 K.A. Bronnikov and V.D. Ivashchuk,
 in:  Reports of 7th Soviet  Conf. on Gravitation
 (ErGU, Erevan, 1988) p. 156-157 [in Russian].

 \bibitem{Tangherlini}
 F.R. Tangherlini, Nuovo Cimento 27 (1963) 636.

 \bibitem{Volovich}
 I.V. Volovich, Class. Quantum Grav.  4 (1987) 183.

 \bibitem{FreundOlson}
 P.G.O. Freund and M. Olson, Phys. Lett. B 199 (1987) 186.

 \bibitem{ADFV}
 I. Ya. Aref'eva, B. Dragovich, P. Frampton and I.V. Volovich,
 Wave function of the universe and p-adic gravity, Steklov
 Mathematical Institute Preprint (1990).

 \bibitem{IMF-90-Yak}
 V.D. Ivashchuk, V.N. Melnikov and S.B. Fadeev, in: Proc.
 National Workshop on Gravitation and Gauge Theory, Yakutsk (1990).

 \bibitem{Mahler}
 K. Mahler, Introduction to p-adic numbers and their functions
(Cambridge Univ. Press., Cambridge, 1973).

 \bibitem{FIM-91-ChPL}
 S.B. Fadeev, V.D. Ivashchuk and V.N. Melnikov, Chinese Phys.
 Lett. (1991) [8 (1991) 439].

\end{thebibliography}
\end{document}